\documentclass[aps,showpacs,nofootinbib]{revtex4}
\usepackage{epsf,latexsym}

\begin{document}

\title{The existence of a minimum wavelength for photons}
\author{Alessandro Pesci}
\email{pesci@bo.infn.it}
\affiliation
{INFN-Bologna, Via Irnerio 46, I-40126 Bologna, Italy}

\begin{abstract}
The holographic property of entropy plays
a key role in the thermodynamic description
of gravitational field equations.
It remains unclear, we argue, whether this property
is necessarily interwoven with gravity itself 
or can be understood instead as a manifestation of
physics outside gravity.

It is pointed out that 
if the latter is the case,
so that the holographic property of entropy 
could be considered as completely rooted
on some microscopic, non-gravitational, intrinsic property of matter,
gravity could merely be understood as the necessary macroscopic manifestation
of this microscopic property of matter, 
being simply the expression of the thermodynamic conservation 
of energy.
A quite peculiar status
of Einstein's gravity in this context
(as compared to generalized metric theories of gravity)
is apparent.
A successful microscopic property of matter,
is found to necessarily mean in particular 
the existence of an absolute minimum wavelength for photons, 
of orders of Planck length.
\end{abstract}


\maketitle


$ $



In the fundamental paper \cite{Jac}, Jacobson showed
that the Einstein equation can be viewed as 
an equation of state.
It should then consist
in a relation among thermodynamic
quantities, describing the thermodynamic limit
of an unknown underlying theory of 
microscopic spacetime degrees of freedom,
i.e. it would be expression of thermodynamics of spacetime.

This being gravity a thermodynamic manifestation of spacetime
has turned out to be not peculiar to Einstein's theory.
The result \cite{Jac}, indeed,
has been shown to hold true for the equations of motion
of generalized theories of gravity \cite{Eli, Bru3}.
An expression/formulation of the action principle, 
with direct thermodynamic meaning, 
has also been found for any diffeomorphism invariant
theory \cite{Pad13}.
Moreover, 
an on-shell equipartition law has been shown to hold,
expressing the energy of a gravitating body
in terms of spacetime degrees of freedom living 
on space 2-surfaces enclosing the source
\cite{Pad7,Pad9}.   

In \cite{Pesci7} the attempt was made to introduce
a complementary thermodynamic perspective,
in which the Einstein equation could also be understood as
merely what must happen in order that a certain intrinsic property
of (matter) entropy be preserved.
This could be intended as an effort to try
to inquire further, within the line opened in \cite{Jac},
as to the meaning of 
having the equation a statistical nature.

The crucial property of matter entropy used in \cite{Pesci7}
is that expressed by the holographic principle \cite{tHo, Sus},
in the formulation through
the generalized covariant entropy bound (GCEB) \cite{Wal5}. 
This property has been suggested from combined consideration
of the existence 
of gravitational collapse
in general relativity 
and of the basic principles of quantum mechanics.
It could seem thus circular to use it
to `derive' Einstein's equation.   
The proposal in \cite{Pesci7}
was, however, 
to consider it as a primeval, or intrinsic, property,
in the sense of something which, being intrinsic to matter,
is supposed to come `before' gravity.
Thus, something whose roots/justification 
are not in gravity.
From this perspective,
the problem of having an understanding of gravity
seems then shifted
to gain insight into the
physical meaning
of a non-gravitational holographic property for entropy.
Aim of the note is to present two remarks upon
this.

The first remark is as follows.
The holographic property just mentioned
is usually formulated
saying that the entropy of a patch of
horizon is proportional to the area,
$S = \eta A$
with $\eta$ a constant, 
and this expression was indeed 
assumed in \cite{Jac}.
%
%
Its motivation in \cite{Jac} was not
the holographic principle, 
but the possible identification of $S$
with the (finite) entanglement entropy
associated to the correlations between the modes
on the two sides of the horizon,
not having, as such, any reference to gravity
(thinking, locally, to the correlations on the two sides
of a Rindler horizon in flat spacetime).
This property of entropy is thus
`intrinsic' 
with the very same meaning as we said for \cite{Pesci7}.
A finite entanglement entropy is obtained
assuming there is a fundamental cutoff length $l_c$
in the modes of the quantum field theory.
The entropy can then be expected to be
$S = \eta A = \alpha A/l_c^2$
for $D = 4$ spacetime dimensions,\footnote{
$S = \eta(D) A = \alpha(D) A/l_c^{D-2}$, in general.}
where the constants $\eta$, $l_c$ and $\alpha$
are intended as not rooted on gravity.\footnote{
The interpretation of (Rindler) horizon entropy
as entanglement entropy has been recently questioned
in \cite{Rov}. There, the suggestion is made to consider it, instead,
as the Shannon entropy coming from Heisenberg uncertainty
associated to the boost generator of accelerated observer.
As the two entropies, even if different in nature,
are equal (and thus with a same functional dependence
on area $A$ of the horizon), 
and both are defined in flat spacetime (and, in this sense, both are
independent of gravity), they seem to share the features
essential for the discussion we are trying to carry on.}

Let us consider
the equilibrium process discussed in \cite{Jac}, 
at a point $P$ of $D = 4$ spacetime.
A certain amount of matter energy $dE$ goes beyond
a local Rindler horizon at $P$, i.e. it is received by the `system'
consisting of the degrees of freedom beyond the horizon.
The variation of entropy of the system
is given by the variation of horizon entropy $dS$,
i.e. by a variation of the area of the horizon, 
geometrically governed by the Raychaudhuri equation,
and this gives a lensing, 
if the first law of thermodynamics 
for this process, $dE = T dS$ ($T$ is horizon temperature),
has to be met.
The strength of the lensing
turns out to be that prescribed
by the Einstein equation
if $\eta = 1/4 G$ (in $c= 1 = \hbar$ units),
where $G$ is the Newton constant.\footnote{
In generalized theories (with $D \geq 4$),
the mentioned results \cite{Eli, Bru3}
show that, 
using 
as horizon entropy
the Noether charge entropy $S_c$ \cite{Wal3},
the thermodynamic relation $dE = T dS_c$ 
is equivalent to the field equations
provided $S_c = (1/4 G_{eff}) A$, 
being $G_{eff}$
the effective gravitational coupling of the theory
and $A$ the horizon area.
Horizon entropy turns out to be, still,
proportional to (actually, a quarter of) horizon area 
if the latter is measured
in units of the effective gravitational coupling \cite{Bru2}.}

The emphasis in \cite{Jac} is that this argument implies 
the Einstein equation can be viewed as
playing the role of an equation of state.
From the perspective of \cite{Pesci7} the same argument
can be read also as follows:
the holographic property of entropy,
i.e. the mere existence of the fundamental cutoff $l_c$
in quantum modes,
determines a lensing of null geodesics
if the first law of thermodynamics, as applied
to the local Rindler horizons, has to be satisfied.\footnote{
A point of view on Jacobson's result
which seems not so different from that presented
in \cite{Lee1}; there, within an information-theoretic perspective,
the stress is, still,
on gravity being derived
from an intrinsic property of entropy.} 
That is to say,
the spacetime of
a world with a fundamental cutoff $l_c$ in quantum field description
is necessarily curved,
if it has in it a local conservation of energy in the thermodynamic sense.
Without a fundamental cutoff,
horizon (entanglement) entropy would be infinite 
and the variations of it (if sensible),
required by the thermodinamic first law,
could be accounted for also without resort to lensing.
In this perspective,
gravity {\it is} the manifestation of 
the existence of $l_c$.

The second remark is similar to the first one. 
It consists in a reformulation of it
in the context of the argument provided in \cite{Pesci7}.
This seemingly adds some indication on
the concrete physical meaning of $l_c$.

In \cite{Pesci7} the holographic property
of entropy is intended to be expressed by
the GCEB.
In particular, the matter entropy content 
$S_m$ of a thin, plane slice of matter
at a given point of spacetime 
is, according to it, bounded from above
by $\Delta A / 4 G$,
being  $\Delta A$ the area variation
of the cross-section 
of the (terminated) lightsheet built on it.  

The presence of $G$ indicates, of course,
the gravitational origin of the bound.
In \cite{Pesci7} however, its existence and value
are supposed to be rooted into physics outside gravity,
and
the gravitational focusing, with strength $G$, 
is supposed to be the necessary consequence of
the universal validity of the bound.
The strength of the focusing is set
by requiring the bound is exactly attained in the most challenging cases.
These are the lightsheets constructed on as-thin-as-possible 
plane layers of matter
--i.e. they coincide with the local Rindler horizons of \cite{Jac}--,
for the most entropic matter.
As apparent in \cite{Pesci7},
to require 
in these circumstances
the validity of the GCEB
is equivalent
to require the validity of first law of thermodynamics 
for local Rindler horizons as in \cite{Jac}
(with $dS_m$ on the lightsheet coinciding
with $\delta Q/T$ of \cite{Jac}).

We try now to proceed further in the approach \cite{Pesci7}
and inspect the possible meaning/justification 
of the GCEB-based holographic 
property of entropy 
in a framework in which
no reference is made to gravity. 
To this end we can make use of the relation
$s \leq \pi \lambda (\rho + p)$
(see \cite{Pesci7} and references therein),
which (we conjectured) is of pure quantum mechanical origin,
and appears to be exactly attained by the most entropic systems,
in particular by the photon gas.
Here $s$, $\rho$ and $p$ are the entropy and energy densities
and pressure respectively, and $\lambda$ is
the quantum wavelength of matter constituents at the
assigned thermodynamic conditions.\footnote{
More precisely, $\lambda$ is defined as that spatial scale $l$
at which the momentum quantum uncertainty 
induced on the constituent particles --if 
imagined physically constrained in $l$-- 
is equal to 
the intrinsic spread in momentum at the assigned
thermodynamic conditions.}

Assuming the bits of information are carried
by the constituent particles so that $\lambda$ is
the spatial scale assigned to $\approx$ 1 bit,
what this relation says is that,
at a given energy + pressure energy in a layer
of 1-bit thickness
there is a maximum number of bits allowed in the layer,
and this maximum number is reached by the photon gas
(the photon gas with the given $\rho + p$).
In a photon gas the distance between the photons
is $\approx \lambda$ so that the layer of photons
can be seen as a layer of bits 
placed side-by-side, i.e. they are maximally packed.
That is,
given a surface with area $A$ 
there is a maximum number of bits
we can put in a layer on that surface,
at the given energy + pressure energy in the layer.
No limit can be envisaged for the number of bits,
if the energy + pressure energy can grow unlimitedly.
All this from quantum mechanics alone.

What the GCEB holographic property of entropy adds,
is that on the area $A$ an absolute maximum number of bits $N_{abs}$
must exist, since
$S_m \leq \Delta A / 4 G \leq A / 4 G 
\equiv N_{abs}$.
The
GCEB holographic property can then be rephrased as:
Given a surface with area A,
there is an absolute maximum number of bits of information 
we can have on a layer of 1-bit thickness on the surface
(i.e. independent of energy and pressure energy in the layer).
We see this amounts to require the existence
of an absolute minimum for the size of a bit
and, equivalently, an absolute minimum for the $\lambda$
of the photons, of orders of the Planck length.

Thus:\\
i) The ultimate physical content of the holographic property
of entropy would be
the existence of an absolute minimum wavelength $l_m$ for photons;\\
ii) The reason of the existence of $l_m$ would not be rooted in gravity;\\
iii) Gravity would be the macroscopic manifestation
of the existence of $l_m$, and would consist in the thermodynamic conservation 
of energy in all circumstances, Rindler observers included.\\
This concludes our second remark.

A final comment.
Changing the functional expression of the entropy assigned to causal horizons
means to change the gravitational equations of motion implied
by the thermodynamic first law.
As mentioned in footnote 3,
there is evidence \cite{Bru2} that
horizon entropy in generalized theories of gravity
is still given by a quarter of the area, 
provided the horizon area
is measured in units
of the effective
gravitational coupling $G_{eff}$.
The identification of horizon entropy with entanglement entropy
is thus still a viable option, with cutoff length $l_c$ (or
the minimum photon wavelength $l_m$ of the photon gas of the second remark)
depending on the theory of gravity (see also \cite{Pad11}). 
The existence of a cutoff length can be suspected 
as the general recipe for a thermodynamic
description of the equations of motion to be feasible.
However, in generalized theories
$l_c$ would have a dependence on curvature,
contrary to Einstein's theory, 
and its value will be set by combined action of
non-gravitational physics and
the gravitational metric theory under examination. 
In the spirit of present note, which aims at ``explaining''
gravity as manifestation of the existence of an absolute 
minimum length fully rooted in non-gravitational physics,
we see that Einstein's theory seems to have a quite unique status.
The Einstein theory (intended as the theory coming from
the Einstein-Hilbert Lagrangian, in $D=4$ or in $D>4$ dimensions) 
appears to be the only metric theory of gravity which allows for a minimum 
cutoff length entirely motivated outside gravity.
This point is crucial in the discussion here,
and is perhaps worth appreciating in general when
comparing the different approaches to gravity. 

I thank Franco Anselmo 
for helpful correspondence 
and Nicol\`o Masi for a fruitful
conversation.


\end{document}